\documentclass[11pt]{article}%
\usepackage{amsfonts}
\usepackage{amsmath}
\usepackage{amssymb}
\usepackage{graphicx}%
\providecommand{\U}[1]{\protect\rule{.1in}{.1in}}
\newtheorem{theorem}{Theorem}
\newtheorem{acknowledgement}[theorem]{Acknowledgement}

\begin{document}

\title{The Symplectic Egg}
\author{Maurice A. de Gosson\thanks{maurice.de.gosson@univie.ac.at}\\University of Vienna\\Faculty of Mathematics, NuHAG\\Nordbergstr. 15, 1090 Vienna}
\maketitle

\begin{abstract}
We invite the reader (presumably an upper level undergraduate student) to a
journey leading from the continent of Classical Mechanics to the new
territories of Quantum Mechanics. We'll be riding the symplectic camel and
have William of Occam as travel companion, so no excess baggage is allowed.
The first part of our trip takes us from the symplectic egg to Gromov's
non-squeezing theorem and its dynamical interpretation. The second part leads
us to a symplectic formulation of the quantum uncertainty principle, which
opens the way to new discoveries.

\end{abstract}

\section*{Prologue}

\begin{quotation}
\textit{What's in a name? That which we call a rose by any other name would
smell as sweet.}

\textit{Romeo and Juliet, Act 2, Scene 2 (W. Shakespeare)}
\end{quotation}

Take an egg --preferably a hard boiled one, and cut it in half along its
middle using a very sharp knife. The surface of section will be roughly
circular and have area $\pi r^{2}$. Next, take a new egg of same size, and cut
it this time along a line joining the egg's tops, again as shown in Fig1. This
time we get an elliptic surface of section with area $\pi R^{2}$ larger than
that of the disk we got previously. So far, so good. But if you now take two
\emph{symplectic eggs}, and do the same thing, then both sections will have
exactly same area! Even \textquotedblleft worse\textquotedblright, no matter
along which plane passing through the center of the egg you cut, you will
always get sections having the same area! This is admittedly a very strange
property, which you probably never have experienced (at least in a direct way)
in everyday life. But what is a symplectic egg? The eggs we are cutting are
metaphors for ellipsoids; an ellipsoid is a round ball that has been deformed
by a linear transformation of space, \textit{i.e.} a transformation preserving
the alignment of three, or more, points. In mathematics such transformations
are represented by matrices. Thus the datum of an ellipsoid is the same thing
as the datum of a ball and of a matrix. What we call a symplectic egg is an
ellipsoid corresponding to the case where the matrix is symplectic (we'll
define the concept in a moment). The reason for which the only symplectic egg
you have seen on your breakfast table is flat --a fried egg!-- is that the
number of rows and columns of a symplectic matrix must always be even. Since
we are unable to visualize things in dimension three or more, the only
symplectic eggs that are accessible to our perception are two dimensional. But
what is a symplectic matrix? In the case of smallest dimension two, a matrix
\begin{equation}
S=%
\begin{pmatrix}
a & b\\
c & d
\end{pmatrix}
\label{s1}%
\end{equation}
is symplectic if it has determinant one:
\begin{equation}
ad-bc=1.\label{adbc}%
\end{equation}
In higher dimensions, 4, 6, 8, etc. there are many more conditions: for
instance 10 if the dimension is 4, 21 if it is 6, and $n(2n+1)$ if it is $2n$.
We will write these conditions explicitly in section \ref{sec11}.

So far, so good. But where do symplectic eggs come from, and what are they
good for? Let me first tell you where symplectic matrices come from. They
initially come from the study of motion of celestial bodies, which is really
rich in mathematical concepts, some of these going back to the observations of
Tycho Brahe, and the work of Galileo Galilei and Johannes Kepler (these were
the \textquotedblleft Giants\textquotedblright\ on the shoulder's of which
Isaac Newton stood). But the notion of symplectic matrix, or more generally
that of symplectic transformation, did really have a long time to wait until
it appeared explicitly and was recognized as a fundamental concept. It was
implicit in the work of Hamilton and Lagrange on classical and celestial
mechanics, until the word \textquotedblleft symplectic\textquotedblright\ was
finally coined by the mathematician Hermann Weyl in his book \emph{The
Classical Groups}, edited in 1939, just before World War II. But still then,
as Ian Stewart \cite{stew} reminds us, it was a rather baffling oddity which
presumably existed for some purpose --but which? It was only later agreed that
the purpose of symplectic transformations is dynamics, that is the study of
\textit{motion}. Let me explain this a little bit more in detail: if we have a
physical system consisting of \textquotedblleft particles\textquotedblright%
\ (sand corns, planets, spacecraft, or quarks) it is economical from both a
notational and computational point of view to describe their motion (that is,
their instantaneous location and velocity) by specifying a phase space vector,
which is a matrix having only one column. For instance, if we are dealing with
one single particle with coordinates $(x,y,z)$ and momentum $(p_{x}%
,p_{y},p_{z})$ (the momentum of a particle is just its velocity multiplied by
its mass $m$) the phase space vector will be the column vector whose entries
are $(x,y,z,p_{x},p_{y},p_{z})$ If we have a large number $N$ of particles
with coordinates $(x_{i},y_{i},z_{i})$ and momenta $(p_{x_{i}},p_{y_{i}%
},p_{z_{i}})$ the phase space vector will be obtained by first writing all the
position coordinates and thereafter the momentum coordinates in corresponding
order, their momenta. These vectors form the phase space of our system of
particles. It turns out that the knowledge of a certain function, the
Hamiltonian (or energy) function, allows us to both predict and retrodict the
motion of our particles; this is done by solving (exactly, or numerically) the
Hamilton equations of motion, which are in the case $n=1$ given by%
\begin{equation}
\frac{dx}{dt}=\frac{\partial H}{\partial p}\text{ \ , \ }\frac{dp}{dt}%
=-\frac{\partial H}{\partial x}. \label{Ham1}%
\end{equation}
Mathematically these equations are just a fancy way to write Newton's second
law $F=ma$. That is, knowing exactly the positions and the momenta at some
initial time, we are able to know what these are going to be at any future
time (we can actually also calculate what they were in the past). The
surprising, and for us very welcome fact is that the transformation which
takes the initial configuration to the final configuration is always a
symplectic transformation! These act on the phase vectors, and once this
action is known, we can determine the future of the whole system of particles,
and this at any moment (mathematicians would say we are in presence of a
\textquotedblleft phase space flow\textquotedblright). The relation between
symplectic transformations and symplectic matrices is that we can associate a
symplectic matrix to every symplectic transformation: it is just the Jacobian
matrix of that transformation. In the simplest cases, for instance when no
external forces act on the particles, these matrices are themselves the
symplectic transformations.

The symplectic egg is a special case a deep mathematical theorem discovered in
1985 by the mathematician Gromov \cite{Gromov}, who won the Abel Prize in 2010
for his discovery (the Abel Prize is the \textquotedblleft
true\textquotedblright\ substitute for the Nobel Prize in mathematics, as
opposed to the Fields medal, which is intended to mathematicians under 40).
Gromov's theorem is nicknamed the \textquotedblleft principle of the
symplectic camel\textquotedblright\ \cite{arnold,FP,stew}, and it tells us
that it impossible to squeeze a symplectic egg through a hole in a plane of
\textquotedblleft conjugate coordinates\textquotedblright\ if its radius is
larger than that of the hole. That one can do that with an ordinary (uncooked)
egg is easy to demonstrate in your kitchen: put it into a cup of vinegar (Coca
Cola will do as well) during 24 hours. You will then be able to squeeze that
egg through the neck of a bottle without any effort!

The marvelous thing with the symplectic egg is that it contains quantum
mechanics in a nutshell... er ... an eggshell! Choose for radius the square
root of Planck's constant $h$ divided by $2\pi$. Then each surface of section
will have radius of $h/2$. In \cite{goletta,physlett,Birk,hileyfest} I have
called such a tiny symplectic egg a \textit{quantum blob}. It is possible
--and in fact quite easy if you know the rules of the game-- to show that this
is equivalent to the uncertainty principle of quantum mechanics. The thing to
remember here is that a classical property (\emph{i.e.} a property involving
usual motions, as that of planets for instance), here symbolized by the
symplectic egg, contains as an imprint quantum mechanics! The analogy between
\textquotedblleft classical\textquotedblright\ and \textquotedblleft
quantum\textquotedblright\ can actually be pushed much further, as I have
shown with Basil Hiley \cite{gohi}. But this, together with the notion of
emergence \cite{gotriad}, is another story.

Some of the ideas presented here are found in our \textit{Physics Reports}
paper \cite{physreps} with F. Luef; they are developed and completed here in a
different way more accessible to a general audience.

\section{Notation and terminology}

Position and moment vectors will be written as column vectors%
\[%
\begin{pmatrix}
x_{1}\\
\vdots\\
x_{n}%
\end{pmatrix}
\text{ \ \ and \ \ }%
\begin{pmatrix}
p_{1}\\
\vdots\\
p_{n}%
\end{pmatrix}
\]
and the corresponding phase vector is thus
\[%
\begin{pmatrix}
x\\
p
\end{pmatrix}
=(x,p)^{T}=(x_{1},...,x_{n};p_{1},...,p_{n})^{T}%
\]
where the superscript $^{T}$ indicates transposition. The integer $n$ is
unspecified; we will call it the number of degrees of freedom. If the vector
$(x,p)^{T}$ denotes the phase vector of a system of $N$ particles, then $n=3N$
and the numbers $x_{1},x_{2},x_{3}$, (resp. $p_{1},p_{2},p_{3}$) can be
identified with the positions $x,y,z$ (resp. the momenta $p_{x},p_{y},p_{z}$)
of the first particle, $x_{4},x_{5},x_{6}$, (resp. $p_{4},p_{5},p_{6}$) with
those of the second particle, and so on. This is not the only possible
convention, but our choice has the advantage of making formulas involving
symplectic matrices particularly simple and tractable. For instance, the
\textquotedblleft standard symplectic matrix\textquotedblright\ is here $J=%
\begin{pmatrix}
0 & I_{\mathrm{d}}\\
-I_{\mathrm{d}} & 0
\end{pmatrix}
$ where $I_{\mathrm{d}}$ is the $n\times n$ identity matrix and $0$ the
$n\times n$ zero matrix. Note that%
\begin{equation}
J^{2}=-I_{\mathrm{d}}\text{ \ , \ }J^{T}=J^{-1}=-J. \label{J}%
\end{equation}

\section{The Symplectic Egg\label{sec1}}

\subsection{Symplectic matrices\label{sec11}}

Let $S$ be a (real) matrix of size $2n$. We say that $S$ is a symplectic
matrix if it satisfies the condition
\begin{equation}
S^{T}JS=J \label{stjs}%
\end{equation}
Clearly the standard symplectic matrix $J$ is itself a symplectic matrix.

Assume that we write the matrix $S$ in block form
\begin{equation}
S=%
\begin{pmatrix}
A & B\\
C & D
\end{pmatrix}
\label{sn}%
\end{equation}
where $A,B,C,D$ are matrices of size $n$. It is a simple exercise in matrix
algebra to show that condition (\ref{stjs}) is equivalent to the the following
constraints on the blocks $A,B,C,D$%
\begin{equation}
A^{T}C=C^{T}A\text{ \ , \ }B^{T}D=D^{T}B\text{ \textit{, and }}A^{T}%
D-C^{T}B=I_{\mathrm{d}}. \label{ABDC}%
\end{equation}
Notice that the two first conditions mean that both $A^{T}C$ and $B^{T}D$
\textit{are symmetric}). Observe that these conditions collapse to
(\ref{adbc}) when $n=1$: in this case $A,B,C,D$ are the numbers $a,b,c,d$ so
that $A^{T}C=ac$ and $B^{T}D=bd$ are automatically symmetric; the condition
$A^{T}D-C^{T}B=I_{\mathrm{d}}$ reduces to $ad-bc=1$.

The product of two symplectic matrices is a symplectic matrix: if $S$ and
$S^{\prime}$ satisfy (\ref{stjs}) then $(SS^{\prime})^{T}JSS^{\prime
}=S^{\prime T}(S^{T}JS)S^{\prime}=S^{\prime T}JS^{\prime}=J$. Also, symplectic
matrices are invertible, and their inverses are symplectic as well: first,
take the determinant of both sides of $S^{T}JS=J$ we get $\det(S^{T}JS)=\det
J$; since $\det J=1$ this is $(\det S)^{2}=1$ hence $S$ is indeed invertible.
Knowing this, we rewrite $S^{T}JS=J$ as $JS=(S^{-1})^{T}J$, from which follows
that $(S^{-1})^{T}JS^{-1}=JSS^{-1}=J$ hence $S^{-1}$ is symplectic. The
symplectic matrices of same size thus form a group, called the symplectic
group and denoted by $\operatorname*{Sp}(2n)$. An interesting property is that
the symplectic group is closed under transposition: if $S$ is a symplectic
matrix, then so is $S^{T}$ (to see this, just take the inverse of the equality
$(S^{-1})^{T}JS^{-1}=J$). Since this means that a matrix is symplectic if and
only if its transpose is, inserting $S^{T}$ in (\ref{stjs}) and noting that
$(S^{T})^{T}=S$ we get the condition
\begin{equation}
SJS^{T}=J\text{.\ } \label{sjst}%
\end{equation}
Replacing $S=%
\begin{pmatrix}
A & B\\
C & D
\end{pmatrix}
$ with $S^{T}=%
\begin{pmatrix}
A^{T} & C^{T}\\
B^{T} & D^{T}%
\end{pmatrix}
$ the conditions (\ref{ABDC}) are thus equivalent to the set of conditions:%
\begin{equation}
AB^{T}=BA^{T}\text{ \ , \ }CD^{T}=DC^{T}\text{ \textit{, }}AD^{T}%
-BC^{T}=I_{\mathrm{d}}. \label{ad}%
\end{equation}
One can obtain other equivalent sets of conditions by using the fact that
$S^{-1}$ and $(S^{-1})^{T}$ are symplectic (see \cite{Birk}).

It is very interesting to note that the inverse of a symplectic matrix is%
\begin{equation}
S^{-1}=%
\begin{pmatrix}
D^{T} & -B^{T}\\
-C^{T} & A^{T}%
\end{pmatrix}
. \label{sinv}%
\end{equation}
It is interesting because this formula is very similar to that giving the
inverse $%
\begin{pmatrix}
d & -b\\
-c & a
\end{pmatrix}
$ of a $2\times2$ matrix $%
\begin{pmatrix}
a & b\\
c & d
\end{pmatrix}
$ with determinant one. The inversion formula (\ref{sinv}) suggests that in a
sense symplectic matrices try very hard to mimic the behavior of $2\times2$
matrices. We will see that this is actually the essence of symplectic
geometry, and at the origin of the symplectic egg property!

A last property of symplectic matrices: recall that when we wanted to show
that a symplectic matrix always is invertible, we established the identity
$(\det S)^{2}=1$. From this follows that the determinant of a symplectic
matrix is a priori either $1$ or $-1$. It turns out --but there is no really
elementary proof of this-- that we always have $\det S=1$ (see for instance
\S 2.1.1 in \cite{Birk} where I give one proof of this property; Mackey and
Mackey's online paper \cite{mack} give a nice discussion of several distinct
methods for proving that symplectic matrices have determinant one.

Conversely, it is not true that any $2n\times2n$ matrix with determinant one
is symplectic when $n>1$. Consider for instance%
\begin{equation}
M=%
\begin{pmatrix}
a & 0 & 0 & 0\\
0 & 1/a & 0 & 0\\
0 & 0 & a & 0\\
0 & 0 & 0 & 1/a
\end{pmatrix}
\label{konter0}%
\end{equation}
where $a\neq0$; this matrix trivially has determinant one, but the condition
$AD^{T}-BC^{T}=I_{\mathrm{d}}$ in (\ref{ad}) is clearly violated unless
$a=\pm1$. Another simple example is provided by%
\[
M=%
\begin{pmatrix}
R(\alpha) & 0\\
0 & R(\beta)
\end{pmatrix}
\]
where $R(\alpha)$ and $R(\beta)$ are rotation matrices with angles $\alpha
\neq\beta$ (this counterexample generalizes to an arbitrary number $2n$ of
phase space dimensions).

\subsection{The first Poincar\'{e} invariant}

In what follows $\gamma(t)$, $0\leq t\leq2\pi$, is a loop in phase space: we
have $\gamma(t)=%
\begin{pmatrix}
x(t)\\
p(t)
\end{pmatrix}
$ where $x(0)=x(2\pi)$, $p(0)=p(2\pi)$; the functions $x(t)$ and $p(t)$ are
supposed to be continuously differentiable. By definition, the first
Poincar\'{e} invariant associated to $\gamma(t)$ is the integral%
\begin{equation}
I(\gamma)=\oint\nolimits_{\gamma}pdx=\int_{0}^{2\pi}p(t)^{T}\dot{x}(t)dt.
\label{firstpoinc}%
\end{equation}
The fundamental property --from which almost everything else in this paper
stems-- is that $I(\gamma)$ is a symplectic invariant. By this we mean that if
we replace the loop $\gamma(t)$ by the a new loop $S\gamma(t)$ where $S$ is a
symplectic matrix, the first Poincar\'{e} invariant will keep the same value:
$I(S\gamma)=I(\gamma)$, that is%
\begin{equation}
\oint\nolimits_{\gamma}pdx=\oint\nolimits_{S\gamma}pdx. \label{pdx}%
\end{equation}
The proof is not very difficult if we carefully use the relations
characterizing symplectic matrices (see Arnol'd \cite{Arnold}, \S 44, p.239,
for a shorter but more abstract proof). We will first need a differentiation
rule for vector-valued functions, generalizing the product formula
$d(uv)/dt=u(dv/dt)+v(du/dt)$ from elementary calculus. Suppose that
\[
u(t)=%
\begin{pmatrix}
u_{1}(t)\\
\vdots\\
u_{n}(t)
\end{pmatrix}
\text{ \ , \ }v(t)=%
\begin{pmatrix}
v_{1}(t)\\
\vdots\\
v_{n}(t)
\end{pmatrix}
\]
are vectors depending on the variable $t$ and such that each component
$u_{j}(t)$, $v_{j}(t)$ is differentiable. Let $M$ be a symmetric matrix of
size $n$ and consider the real-valued function $u(t)^{T}Mv(t)$. That function
is differentiable as well and its derivative is given by the formula
\begin{equation}
\frac{d}{dt}\left[  u(t)^{T}Mv(t)\right]  =\dot{u}(t)^{T}Mv(t)+u(t)^{T}%
M\dot{v}(t) \label{diff}%
\end{equation}
(we are writing $\dot{u},\dot{v}$ for $du/dt$, $dv/dt$ as is customary in
mechanics); for a proof I refer you to your favorite calculus book.

Let us now go back to the proof of the symplectic invariance of the first
Poincar\'{e} invariant. We write as usual the symplectic matrix $S$ in block
form $%
\begin{pmatrix}
A & B\\
C & D
\end{pmatrix}
$ so that the loop $S\gamma(t)$ is parametrized by%
\[
S\gamma(t)=%
\begin{pmatrix}
Ax(t)+Bp(t)\\
Cx(t)+Dp(t)
\end{pmatrix}
\text{ \ , \ }0\leq t\leq2\pi.
\]
We thus have, by definition of the Poincar\'{e} invariant,
\[
I(S\gamma)=\int_{0}^{2\pi}(Cx(t)+Dp(t))^{T}(A\dot{x}(t)+B\dot{p}(t))dt;
\]
expanding the product in the integrand, we have $I(S\gamma)=I_{1}+I_{2}$ where%
\begin{align*}
I_{1}  &  =\int_{0}^{2\pi}x(t)^{T}C^{T}A\dot{x}(t)dt+\int_{0}^{2\pi}%
p(t)^{T}D^{T}B\dot{p}(t)dt\\
I_{2}  &  =\int_{0}^{2\pi}x(t)^{T}C^{T}B\dot{p}(t)dt+\int_{0}^{2\pi}%
p(t)^{T}D^{T}A\dot{x}(t)dt.
\end{align*}
We claim that $I_{1}=0$. Recall that $C^{T}A$ and $C^{T}B$ are symmetric in
view of the two first equalities in (\ref{ABDC}); applying the differentiation
formula (\ref{diff}) with $u=v=x$ we have%
\begin{align*}
\int_{0}^{2\pi}x(t)^{T}C^{T}A\dot{x}(t)dt  &  =\frac{1}{2}\int_{0}^{2\pi}%
\frac{d}{dt}(x(t)^{T}C^{T}Ax(t))dt\\
&  =\frac{1}{2}\left[  x(2\pi)C^{T}Ax(2\pi)-x(0)C^{T}Ax(0)\right] \\
&  =0
\end{align*}
because $x(0)=x(2\pi)$. Likewise, applying (\ref{diff}) with $u=v=p$ we get%
\[
\int_{0}^{2\pi}p(t)D^{T}B\dot{p}(t)dt=0
\]
hence $I_{1}=0$ as claimed. We next consider the term $I_{2}$. Rewriting the
integrand of the second integral as
\[
x(t)^{T}C^{T}B\dot{p}(t)=\dot{p}(t)^{T}B^{T}Cx(t)^{T}%
\]
(because it is a number, and hence equal to its own transpose!) we have%
\[
I_{2}=\int_{0}^{2\pi}\dot{p}(t)^{T}B^{T}Cx(t)^{T}dt+\int_{0}^{2\pi}%
p(t)^{T}D^{T}A\dot{x}(t)dt
\]
that is, since $D^{T}A=I_{\mathrm{d}}+B^{T}C$ by transposition of the third
equality in (\ref{ABDC}),
\[
I_{2}=\int_{0}^{2\pi}p(t)^{T}\dot{x}(t)dt+\int_{0}^{2\pi}\left[  p(t)^{T}%
B^{T}CA\dot{x}(t)+\dot{p}(t)^{T}B^{T}CAx(t)\right]  dt.
\]
Using again the rule (\ref{diff}) and noting that the first integral is
precisely $I(\gamma)$ we get, $D^{T}A$ being symmetric,%
\[
I_{2}=I(\gamma)+\int_{0}^{2\pi}\frac{d}{dt}\left[  p(t)^{T}B^{T}CAx(t)\right]
dt.
\]
The equality $I(S\gamma)=I(\gamma)$ follows noting that the integral in the
right-hand side is
\[
p(2\pi)^{T}B^{T}CAx(2\pi)-p(0)^{T}B^{T}CAx(0)=0
\]
since $(x(2\pi),p(2\pi))=(x(0),p(0))$.

The observant reader will have observed that we really needed all of the
properties of a symplectic matrix contained in the set of conditions
(\ref{ABDC}); this shows that the symplectic invariance of the first
Poincar\'{e} invariant is a characteristic property of symplectic matrices.

\subsection{Proof of the symplectic egg property\label{subsec3}}

Let us denote by $B_{R}$ the phase space ball centered at the origin and
having radius $R$. It is the set of all points $z=(x,p)$ such that
$|z|^{2}=|x|^{2}+|p|^{2}\leq R^{2}$. What we call a \textquotedblleft
symplectic egg\textquotedblright\ is the image $S(B_{R})$ of $B_{R}$ by a
symplectic matrix $S$. It is thus an ellipsoid in phase space, consisting of
all points $z$ such that $S^{-1}z$ is in the ball $B_{R}$, that is
$|S^{-1}z|^{2}\leq R^{2}$. Using formula (\ref{sinv}) giving the inverse of
$S=%
\begin{pmatrix}
A & B\\
C & D
\end{pmatrix}
$ together with the relations $A^{T}C=C^{T}A$, $B^{T}D=D^{T}B$ in (\ref{ABDC})
we get the following explicit expression after some easy calculations:%
\[
x^{T}(CC^{T}+DD^{T})x-2x^{T}(DB^{T}+CA^{T})p+p^{T}(AA^{T}+BB^{T})p\leq R^{2}%
\]
(don't worry: we will not have to use this cumbersome inequality in what follows!).

Let us now cut $S(B_{R})$ by a plane $\Pi_{j}$ of conjugate coordinates
$x_{j},p_{j}$. We get an elliptic surface $\Gamma_{j}$, whose boundary is an
ellipse denoted by $\gamma_{j}$. Since that ellipse lies in the plane $\Pi
_{j}$ we can parametrize it by only specifying coordinates $x_{j}(t)$,
$p_{j}(t)$ all the other being identically zero; relabeling if necessary the
coordinates we may as well assume that $j=1$ so that the curve $\gamma_{j}$
can be parametrized as follows:
\[
\gamma_{j}(t)=(x_{1}(t),0,\cdot\cdot\cdot,0;p_{1}(t),0,\cdot\cdot\cdot,0)^{T}%
\]
for $0\leq t\leq2\pi$ with $x_{1}(0)=x_{1}(2\pi)$ and $p_{1}(0)=p_{1}(2\pi)$.
Since $x_{k}(t)=0$ and $p_{k}(t)=0$ for $k>1$ the area of the ellipse is given
by the formula%
\begin{align*}
\operatorname*{Area}(\Gamma_{1})  &  =\int_{0}^{2\pi}p_{1}(t)\dot{x}%
_{1}(t)dt\\
&  =\sum_{k=1}^{n}\int_{0}^{2\pi}p_{k}(t)\dot{x}_{k}(t)dt\\
&  =\oint\nolimits_{\gamma_{1}}pdx
\end{align*}
hence $\operatorname*{Area}(\Gamma_{1})=I(\gamma_{1})$. Since the inverse
matrix $S^{-1}$ is symplectic, we have $I(\gamma_{1})=I(S^{-1}\gamma_{1})$.
But the loop $S^{-1}\gamma_{1}$ bounds a section of the ball $B_{R}$ by a
plane (the plane $S^{-1}\Pi_{j}$) passing through its center. This loop is
thus a great circle of $B_{R}$ and the area of the surface $S^{-1}\Gamma_{1}$
is thus exactly $\pi R^{2}$, which was to be proven.

We urge the reader to notice that the assumption that we are cutting
$S(B_{R})$ with a plane of \textit{conjugate }coordinates\textit{ }is
\emph{essential}, because it is this assumption that allowed us to identify
the area of the section with action. Here is a counterexample which shows that
the property does not hold for arbitrary sections of $S(B_{R}).$ Take, for
instance%
\begin{equation}
S=%
\begin{pmatrix}
\lambda_{1} & 0 & 0 & 0\\
0 & \lambda_{2} & 0 & 0\\
0 & 0 & 1/\lambda_{1} & 0\\
0 & 0 & 0 & 1/\lambda_{2}%
\end{pmatrix}
\text{ \ \ , \ }\lambda_{1}>0\text{, }\lambda_{2}>0,\text{ and }\lambda
_{1}\neq\lambda_{2} \label{konter}%
\end{equation}
so that $S(B_{R})$ is defined by the inequality%
\[
\frac{1}{\lambda_{1}}x_{1}^{2}+\frac{1}{\lambda_{2}}x_{2}^{2}+\lambda_{1}%
p_{1}^{2}+\lambda_{2}p_{2}^{2}\leq R^{2}.
\]
The section of $S(B_{R})$ by the $x_{2},p_{2}$ plane is the ellipse%
\[
\frac{1}{\lambda_{1}}x_{1}^{2}+\lambda_{1}p_{1}^{2}\leq R^{2}%
\]
which has area $\pi(R^{2}\sqrt{\lambda_{1}}\sqrt{1/\lambda_{1}})=\pi R^{2}$ as
predicted, but its section with the $x_{2},p_{1}$ plane is the ellipse%
\[
\frac{1}{\lambda_{1}}x_{1}^{2}+\lambda_{2}p_{2}^{2}\leq R^{2}%
\]
which has area $\pi(R^{2}\sqrt{\lambda_{1}/\lambda_{2}})$ which is different
from $\pi R^{2}$ since $\lambda_{1}\neq\lambda_{2}$.

The assumption that $S$ is symplectic is also essential. Assume that we
scramble the diagonal entries of the matrix $S$ above in the following way:%
\[
S^{\prime}=%
\begin{pmatrix}
\lambda_{1} & 0 & 0 & 0\\
0 & \lambda_{2} & 0 & 0\\
0 & 0 & 1/\lambda_{2} & 0\\
0 & 0 & 0 & 1/\lambda_{1}%
\end{pmatrix}
\text{.}%
\]
The matrix $S^{\prime}$ still has determinant one, but it is not symplectic
(cf. (\ref{konter0})). The section $S^{\prime}(B_{R})$ by the $x_{2},p_{2}$
plane is now the ellipse
\[
\frac{1}{\lambda_{1}}x_{1}^{2}+\lambda_{2}p_{1}^{2}\leq R^{2}%
\]
with area $\pi R^{2}\sqrt{\lambda_{1}/\lambda_{2}}\neq\pi R^{2}.$

\section{The Symplectic Camel}

The property of the symplectic camel is a generalization of the property of
the symplectic egg for arbitrary canonical transformations; it reduces to the
latter in the linear case.

\subsection{Gromov's non-squeezing theorem: static formulation}

As we mentioned in the Prologue, the property of the symplectic egg is related
to the \textquotedblleft non-squeezing theorem\textquotedblright\ of Gromov
\cite{Gromov} in 1985. To understand it fully we have to introduce the notion
of canonical transformation \cite{Arnold,Goldstein}. A canonical
transformation is an invertible infinitely differentiable mapping%
\[
f:%
\begin{pmatrix}
x\\
p
\end{pmatrix}
\longrightarrow%
\begin{pmatrix}
x^{\prime}\\
p^{\prime}%
\end{pmatrix}
\]
of phase space on itself whose inverse $f^{-1}$ is also infinitely
differentiable and such that its Jacobian matrix
\[
f^{\prime}(x,p)=\frac{\partial(x^{\prime},p^{\prime})}{\partial(x,p)}%
\]
is symplectic at every point $(x,p)$. A symplectic matrix $S=%
\begin{pmatrix}
A & B\\
C & D
\end{pmatrix}
$ automatically generates a linear canonical transformation by letting it act
on phase space vectors $%
\begin{pmatrix}
x\\
p
\end{pmatrix}
\longrightarrow S%
\begin{pmatrix}
x\\
p
\end{pmatrix}
$: it is an invertible transformation (because symplectic matrices are
invertible), trivially infinitely differentiable, and the Jacobian matrix is
the matrix $S$ itself. Phase space translations, that is mappings $%
\begin{pmatrix}
x\\
p
\end{pmatrix}
\longrightarrow%
\begin{pmatrix}
x+x_{0}\\
p+p_{0}%
\end{pmatrix}
$ are also canonical: their Jacobian matrix is just the identity $%
\begin{pmatrix}
I_{\mathrm{d}} & 0\\
0 & I_{\mathrm{d}}%
\end{pmatrix}
$. By composing linear canonical transformations and translations one obtains
the class of all affine canonical transformations.

Here is an example of a nonlinear canonical transformation: assume that $n=1$
and denote the phase space variables by $r$ and $\varphi$ instead of $x$ and
$p$; the transformation defined by $(r,\varphi)\longrightarrow(x,p)$ with%
\[
x=\sqrt{2r}\cos\varphi\text{ \ , \ }p=\sqrt{2r}\sin\varphi\text{ \ , \ }%
0\leq\varphi<2\pi,
\]
has Jacobian matrix%
\[
f^{\prime}(r,\varphi)=%
\begin{pmatrix}
\frac{1}{\sqrt{2r}}\cos\varphi & \frac{1}{\sqrt{2r}}\sin\varphi\\
-\sqrt{2r}\sin\varphi & \sqrt{2r}\cos\varphi
\end{pmatrix}
\]
which has determinant one for every choice of $r$ and $\varphi$. The
transformation $f$ is thus canonical, and can be extended without difficulty
to the multi-dimensional case by associating a similar transformation to each
pair $(x_{j},p_{j})$. It is in fact a symplectic version of the usual passage
to polar coordinates (the reader can verify that the latter is not canonical
by calculating its Jacobian matrix); it can also be viewed as the simplest
example of action-angle variable \cite{Arnold,Goldstein}; for instance it
reduces the isotropic harmonic oscillator Hamiltonian $H=\frac{1}{2}%
(p^{2}+x^{2})$ to $K=r$.

We will see in a moment why canonical transformations play such an important
role in Physics (and especially in classical mechanics), but let us first
state Gromov's theorem:

\begin{description}
\item[Gromov's theorem:] \emph{No canonical transformation can squeeze a ball
}$B_{R}$\emph{ through a circular hole in a plane }$\Pi_{j}$ \emph{of
conjugate coordinates }$x_{j},p_{j}$ \emph{with smaller radius }$r<R$\emph{ }.
\end{description}

This statement is surprisingly simple, and one can wonder why it took so long
time to discover it. There are many possible answers. The most obvious is that
all known proofs Gromov's theorem are extremely difficult, and make use of
highly non-trivial techniques from various parts of pure mathematics, so the
result cannot be easily derived from elementary principles. Another reason is
that it seems, as we will discuss below, to contradict the common conception
of Liouville's theorem, and was therefore unsuspected!

So, what is the relation of Gromov's theorem with our symplectic eggs, and
where does its nickname \textquotedblleft principle of the symplectic
camel\textquotedblright\ come from? The denomination apparently appeared for
the first time in Arnol'd's paper \cite{arnold}. Recalling that in
\cite{Matthew} it is stated that

\begin{quote}
`...\emph{Then Jesus said to his disciples, `Amen, I say to you, it will be
hard for one who is rich to enter the kingdom of heaven. Again I say to you,
it is easier for a camel to pass through the eye of a needle than for one who
is rich to enter the kingdom of God}'.
\end{quote}

\noindent The biblical camel is here the ball $B_{R}$, and the eye of the
needle is the hole in the $x_{j},p_{j}$ plane! (For alternative
interpretations of the word \textquotedblleft camel\textquotedblright; see the
reader's comments following E. Samuel Reich's New Scientist paper \cite{Reich}
about \cite{FP}.)

Let us next show that the section property of the symplectic egg is indeed a
linear (or affine) version of Gromov's theorem. It is equivalent to prove that
no symplectic egg $S(B_{R})$ with radius $R$ larger than that, $r$, of the
hole in the $x_{j},p_{j}$ plane can be threaded through that hole. Passing
$S(B_{R})$ through the hole means that the section of the symplectic egg by
the $x_{j},p_{j}$ plane, which has area $\pi R^{2}$, is smaller than the area
$\pi r^{2}$ of the hole; hence we must have $R\leq r$.

\subsection{Dynamical interpretation}

The reason for which canonical transformations play an essential role in
Physics comes from the fact that Hamiltonian phase flows precisely consist of
canonical transformations. Consider a particle with mass $m$ moving along the
$x$-axis under the action of a scalar potential $V$. The particle is subject
to a force $F=-\frac{d}{dx}V(x)$. Since $F=mdv/dt=dp/dt$ (Newton's second
law), the equations of motion can be written%
\begin{equation}
m\frac{dx}{dt}=p\text{ \ , \ }\frac{dp}{dt}=-\frac{dV}{dx}\text{.}
\label{motion1}%
\end{equation}
In terms of the Hamilton function
\[
H(x,p)=\frac{1}{2m}p^{2}+V(x)
\]
this system of differential equations is equivalent to Hamilton's equations of
motion%
\begin{equation}
\frac{dx}{dt}=\frac{\partial H}{\partial p}\text{ \ , \ }\frac{dp}{dt}%
=-\frac{\partial H}{\partial p}. \label{motion2}%
\end{equation}
We will more generally consider the $n$-dimensional version of (\ref{motion2})
which reads%
\begin{equation}
\frac{dx_{j}}{dt}=\frac{\partial H}{\partial p_{j}}\text{ \ , \ }\frac{dp_{j}%
}{dt}=-\frac{\partial H}{\partial x_{j}}\text{ \ , \ }1\leq j\leq n.
\label{motion3}%
\end{equation}
(In mathematical treatments of Hamilton's equations
\cite{Arnold,Goldstein,Birk} the function $H$ can be of a very general type,
and even depend on time $t$). In either case, these equations determine --as
any system of differential equations does-- a \emph{flow}. By definition, the
Hamiltonian flow is the infinite set of mappings $\phi_{t}^{H}$ defined as
follows: suppose we solve the system (\ref{motion3}) after having chosen
initial conditions $x_{1}(0),...,x_{n}(0)\ $and $p_{1}(0),...,p_{n}(0)$ at
time $t=0$ for the position and momentum coordinates. Denote the initial
vector thus defined $%
\begin{pmatrix}
x(0)\\
p(0)
\end{pmatrix}
$. Assuming that the solution to Hamilton's equations at time $t$ exists (and
is unique), we denote it by $%
\begin{pmatrix}
x(t)\\
p(t)
\end{pmatrix}
$. By definition, $\phi_{t}^{H}$ is just the mapping that takes the initial
vector to the final vector:%
\begin{equation}%
\begin{pmatrix}
x(t)\\
p(t)
\end{pmatrix}
=\phi_{t}^{H}%
\begin{pmatrix}
x(0)\\
p(0)
\end{pmatrix}
. \label{fi}%
\end{equation}
As time varies, the initial point describes a curve in phase space; it is
called a \textquotedblleft flow curve\textquotedblright\ or a
\textquotedblleft Hamiltonian trajectory\textquotedblright.

The essential property to remember is that each mapping $\phi_{t}^{H}$ is a
canonical transformation;\ Hamiltonian flows are therefore volume preserving:
this is Liouville's theorem \cite{Arnold,Goldstein}. This easily follows from
the fact that symplectic matrices have determinant one. Since it is not true
that every matrix with determinant one is symplectic, as soon as $n>1$ volume
preservation also holds for other transformations, and is therefore not a
characteristic property of Hamiltonian flows; see Arnold \cite{Arnold}, Ch.3,
\S 16 for a discussion of this fact. The thing to observe is that volume
preservation does not imply conservation of shape, and one could therefore
imagine that under the action of a Hamiltonian flow a subset of phase space
can be stretched in all directions, and eventually get very thinly spread out
over huge regions of phase space, so that the projections on any plane could
\textit{a priori} become arbitrary small after some (admittedly, very long)
time $t$. In addition, one may very well envisage that the larger the number
$n$ of degrees of freedom, the more that spreading will occur since there are
more directions in which the ball is likely to spread! This possibility, which
is ruled out by the symplectic camel as we will explain below, has led to many
philosophical speculations about Hamiltonian systems. For instance, in his
1989 book Roger Penrose (\cite{penrose}, p.174--184) comes to the conclusion
that phase space spreading suggests that \textquotedblleft\textit{classical
mechanics cannot actually be true of our world}\textquotedblright\ (p.183,
l.--3). In fact, our discussion of Gromov's theorem shows that Hamiltonian
evolution is much less disorderly than Penrose thought. Consider again our
phase space ball $B_{R}$. Its orthogonal projection (or \textquotedblleft
shadow\textquotedblright) on any two-dimensional subspace $\Pi$ of phase space
is a circular surface with area $\pi R^{2}$. Suppose now that we move the ball
$B_{R}$ using a Hamiltonian flow $\phi_{t}^{H}$ and choose for $\Pi$ the plane
$\Pi_{j}$ of conjugate coordinates $x_{j},p_{j}$. The ball will slowly get
deformed, while keeping same volume. But, as a consequence of the principle of
the symplectic camel, its \textquotedblleft shadow\textquotedblright\ on any
plane $\Pi_{j}$ will never decrease below its original value $\pi R^{2}$! Why
is it so? First, it is clear that if the area of the projection of $f(B_{R})$
on on a plane $x_{j},p_{j}$ ($f$ a canonical transformation) will never be
smaller than $\pi R^{2}$, then we cannot expect that $f(B_{R})$ lies inside a
cylinder $(p_{j}-a_{j})^{2}+(x_{j}-b_{j})^{2}=r^{2}$ if $r<R$. So is the
\textquotedblleft principle of the symplectic camel\textquotedblright%
\ stronger than Gromov's theorem? Not at all, it is equivalent to it! Let us
prove this. We assume as in section \ref{subsec3} that $j=1$; this does not
restrict the generality of the argument. Let $\gamma_{1}$ be the boundary of
the projection of $f(B_{R})$ on the $x_{1},p_{1}$ plane; it is a loop
encircling a surface $\Gamma_{1}$ with area at least $\pi R^{2}$. That surface
$\Gamma_{1}$ can be deformed into a circle with same area using an
area-preserving mapping of the $x_{1},p_{1}$ plane; call that mapping $f_{1}$
and define a global phase space transformation $f$ by the formula
\[
f(x_{1},p_{1},x_{2},p_{2},.....,x_{n},p_{n})=(f_{1}(x_{1},p_{1}),x_{2}%
,p_{2},...,x_{n},p_{n})\text{.}%
\]
Calculating the Jacobian matrix it is easy to check that the matrix $f$ is a
canonical transformation, hence our claim. For a more detailed discussion of
this and related topics see \cite{FP,physreps}.

\section{Quantum Blobs}

By definition, a quantum blob is a symplectic egg with radius $R=\sqrt{\hbar
}.$ The section of quantum blob by a plane of conjugate coordinates is thus
$\pi\hbar=\frac{1}{2}h$. We will see that quantum blobs qualify as the
smallest units of phase space allowed by the uncertainty principle of quantum
mechanics. We begin with a very simple example illustrating the basic idea,
which is that a closed (phase space) trajectory cannot be carried by an energy
shell smaller (in a sense to be made precise) than a quantum blob. As simple
as this example is, it allows us to recover the ground energy of the
anisotropic quantum harmonic oscillator.

\subsection{The harmonic oscillator}

The fact that the ground energy level of a one-dimensional harmonic
oscillator
\[
H=\frac{p_{x}^{2}}{2m}+\frac{1}{2}m\omega^{2}x^{2}%
\]
is different from zero is heuristically justified in the physical literature
by the following observation: since Heisenberg's uncertainty relation $\Delta
p_{x}\Delta x\geq\frac{1}{2}\hbar$ prevent us from assigning simultaneously a
precise value to both position and momentum, the oscillator cannot be at rest.
To show that the lowest energy has the value $\frac{1}{2}\hbar\omega$
predicted by quantum mechanics one can then argue as follows: since we cannot
distinguish the origin $(x=0,p=0)$ of phase space from a phase plane
trajectory lying inside the double hyperbola $p_{x}x<\frac{1}{2}\hbar$ , we
must require that the points $(x,p)$ of that trajectory are such that
$|p_{x}x|\geq\frac{1}{2}\hbar$; multiplying both sides of the trivial
inequality
\[
\frac{p_{x}^{2}}{m\omega}+m\omega x^{2}\geq2|px|\geq\hbar
\]
by $\omega/2$ we then get
\[
E=\frac{p_{x}^{2}}{2m}+\frac{1}{2}m\omega^{2}x^{2}\geq\frac{1}{2}\hbar\omega
\]
which is the correct lower bound for the quantum energy. This argument can be
reversed: since the lowest energy of an oscillator with frequency $\omega$ and
mass $m$ is $\frac{1}{2}\hbar\omega$, the minimal phase space trajectory will
be the ellipse
\[
\frac{p_{x}^{2}}{m\hbar\omega}+\frac{x^{2}}{(\hbar/m\omega)}=1
\]
which encloses a surface with area $\frac{1}{2}h$. Everything in this
discussion immediately extends to the generalized anisotropic $n$-dimensional
oscillator
\[
H=\sum_{j=1}^{n}\frac{p_{j}^{2}}{2m_{j}}+\frac{1}{2}m_{j}\omega_{j}^{2}x^{2}%
\]
and one concludes that the smallest possible trajectories in $x_{j},p_{j}$
space are the ellipses%
\[
\frac{p_{j}^{2}}{m_{j}\hbar\omega_{j}}+\frac{x_{j}^{2}}{(\hbar/m_{j}\omega
_{j})}=1\text{.}%
\]
By the same argument as above, using each of the Heisenberg uncertainty
relations
\begin{equation}
\Delta p_{j}\Delta x_{j}\geq\frac{1}{2}\hbar\label{hup}%
\end{equation}
we recover the correct ground energy level%
\[
E=\frac{1}{2}\hbar\omega_{1}+\frac{1}{2}\hbar\omega_{2}+\cdot\cdot\cdot
+\frac{1}{2}\hbar\omega_{n}%
\]
as predicted by standard quantum theory \cite{Messiah}. In addition, one finds
that, the projection of the motion on any plane of conjugate variables
$x_{j},p_{j}$ will always enclose a surface having an area at least equal to
$\frac{1}{2}h$. In other words, the motions corresponding to the lowest
possible energy must lie on a quantum blob!

\subsection{Quantum blobs and uncertainty}

The Heisenberg inequalities (\ref{hup}) are a weak form of the quantum
uncertainty principle; they are a particular case of the more accurate
Robertson--Schr\"{o}dinger \cite{Rob,Schr} inequalities%
\begin{equation}
(\Delta p_{j})^{2}(\Delta x_{j})^{2}\geq\Delta(x_{j},p_{j})^{2}+\tfrac{1}%
{4}\hbar^{2} \label{robup}%
\end{equation}
(see Messiah \cite{Messiah} for a simple derivation). Here, in addition to the
standard deviations $\Delta x_{j}$, $\Delta p_{j}$ we have the covariances
$\Delta(x_{j},p_{j})$ which are a measurement of how much the two variables
$x_{j},p_{j}$ change together. (We take the opportunity to note that the
interpretation of quantum uncertainty in terms of standard deviations goes
back to Kennard \cite{kennard};\ Heisenberg's \cite{heisenberg} own
interpretation was much more heuristic). Contrarily to what is often believed
the Heisenberg inequalities (\ref{hup}) and the Robertson--Schr\"{o}dinger
inequalities (\ref{robup}) are not statements about the accuracy of our
measurements; their derivation assumes on the contrary perfect instruments;
see the discussion in Peres \cite{Peres}, p.93. Their meaning is that if the
same preparation procedure is repeated a large number of times on an ensemble
of systems, and is followed by either by a measurement of $x_{j}$, or by a
measurement of $p_{j}$, then the results obtained have standard deviations
$\Delta x_{j}$, $\Delta p_{j}$; in addition these measurements need not be
independent: this is expressed by the statistical covariances $\Delta
(x_{j},p_{j})$ appearing in the inequalities (\ref{robup}).

It turns out that quantum blobs can be used to give a purely geometric and
intuitive idea of quantum uncertainty. Let us first consider the case $n=1$,
and define the covariance matrix by
\begin{equation}
\Sigma=%
\begin{pmatrix}
\Delta x^{2} & \Delta(x,p)\\
\Delta(p,x) & \Delta p^{2}%
\end{pmatrix}
. \label{covma1}%
\end{equation}
Its determinant is $\det\Sigma=(\Delta p)^{2}(\Delta x)^{2}-\Delta(x,p)^{2}$,
so in this case the Robertson--Schr\"{o}dinger inequality is the same thing as
$\det\Sigma\geq\tfrac{1}{4}\hbar^{2}$. Now to the geometric interpretation. In
statistics it is customary to associate to $\Sigma$ the so-called covariance
ellipse: it is the set of $\Omega_{\Sigma}$ points $(x,p)$ in the phase plane
satisfying
\begin{equation}
\frac{1}{2}(x,p)\Sigma^{-1}%
\begin{pmatrix}
x\\
p
\end{pmatrix}
\leq1. \label{covell1}%
\end{equation}
Its area is $2\pi\sqrt{\det\Sigma}$, that is%
\[
\operatorname{Area}(\Omega_{\Sigma})=2\pi\left[  (\Delta p)^{2}(\Delta
x)^{2}-\Delta(x,p)^{2}\right]  ^{1/2}%
\]
and the inequality $\det\Sigma\geq\tfrac{1}{4}\hbar^{2}$ is thus equivalent to
$\operatorname{Area}(\Omega_{\Sigma})\geq\pi\hbar=\frac{1}{2}h$. We have thus
succeeded in expressing the rather complicated Robertson--Schr\"{o}dinger
inequality (\ref{robup}) in terms of the area of a certain ellipse. In higher
dimensions the same argument applies, but contrarily to what common intuition
suggests, the Robertson--Schr\"{o}dinger inequalities are not expressed in
terms of volume (which is the generalization of area to higher dimensions),
but again in terms of \emph{areas }--namely those of the intersections of the
conjugate planes $x_{j},p_{j}$ with the covariance ellipsoid%
\begin{equation}
\Sigma=%
\begin{pmatrix}
\Delta(x,x) & \Delta(x,p)\\
\Delta(p,x) & \Delta(p,p)
\end{pmatrix}
. \label{covma2}%
\end{equation}
Here $\Delta(x,x),\Delta(x,p)$, etc. are the $n\times n$ block-matrices
$\left(  \Delta(x_{i},x_{j})\right)  _{1\leq i,j\leq n}$, $\left(
\Delta(x_{i},p_{j})\right)  _{1\leq i,j\leq n}$ etc. Notice that the diagonal
terms of $\Sigma$ are just the variances $\Delta x_{1}^{2},...,\Delta
x_{n}^{2};\Delta p_{1}^{2},...,\Delta p_{n}^{2}$ so that (\ref{covma2})
reduces to (\ref{covma1}) for $n=1$. Defining the covariance ellipsoid
$\Omega_{\Sigma}$ as above, one then proves that the inequalities
(\ref{robup}) are equivalent to the property that the intersection of
$\Omega_{\Sigma}$ with the planes $x_{j},p_{j}$ is at least $\frac{1}{2}h$.
These inequalities are saturated (\textit{i.e.} they become equalities) if and
only if these intersections have exactly area $\frac{1}{2}h$, that is, if and
only if $\Omega_{\Sigma}$ is a quantum blob! The proof goes as follows (for a
detailed argument see \cite{FP,physreps}): one first remarks, using a simple
algebraic argument that the Robertson--Schr\"{o}dinger inequalities are
equivalent to the following condition of the covariance matrix, well-known in
quantum optics, see \textit{e.g.} \cite{SMD,SSM} and the references therein:

\begin{quotation}
\textit{The eigenvalues of the Hermitian matrix} $\Sigma+\frac{i\hbar}{2}J$
\textit{are non-negative}: $\Sigma+\frac{i\hbar}{2}J\geq0$.
\end{quotation}

\noindent The next step consists in noting that in view of Sylvester's theorem
from linear algebra that the leading principal minors of
\[
\Sigma+\frac{i\hbar}{2}J=%
\begin{pmatrix}
\Delta(x,x) & \Delta(x,p)+\frac{i\hbar}{2}I\\
\Delta(p,x)-\frac{i\hbar}{2}I & \Delta(p,p)
\end{pmatrix}
\]
are non-negative. This applies in particular to the minors of order $2$ so
that we must have%
\[%
\begin{vmatrix}
\Delta x_{j}^{2} & \Delta(x_{j},p_{j})+\frac{i\hbar}{2}\\
\Delta(p_{j},x_{j})-\frac{i\hbar}{2} & \Delta p_{j}^{2}%
\end{vmatrix}
\geq0
\]
and this condition is precisely the Robertson--Schr\"{o}dinger inequality
(\ref{robup}).

As we have seen, the fact that the covariance ellipsoid is cut by the
conjugate coordinate planes along ellipsoids with areas $\geq\frac{1}{2}h$
implies the Robertson--Schr\"{o}dinger inequalities. This is thus a geometric
statement --and a strong one-- of the quantum uncertainty principle, which can
be rephrased as follows:

\begin{quotation}
\textit{Every quantum covariance ellipsoid contains a quantum blob, i.e. a
symplectic egg with radius} $\sqrt{\hbar}$. \textit{When this ellipsoid is a
quantum blob, the Robertson--Schr\"{o}dinger inequalities are saturated.}
\end{quotation}

This statement can be extended in various ways; in a very recent paper
\cite{sat} we have applied this geometric approach to the quantum uncertainty
principle to the study of partial saturation of the Robertson--Schr\"{o}dinger
inequalities for mixed quantum states. We show, in particular, that partial
saturation corresponds to the case where some (but not all) planes of
conjugate coordinates cut the covariance ellipsoid along an ellipse with
exactly area $\frac{1}{2}h$; this allows us to characterize those states for
which this occurs (they are generalized Gaussians).

Another important thing we will unfortunately not be able to discuss in detail
because of length limitations, is the following: everything we have said above
still holds true if we replace the sentence \textquotedblleft planes of
conjugate coordinates $x_{j},p_{j}$\textquotedblright\ with the sentence
\textquotedblleft symplectic planes\textquotedblright. A symplectic plane is a
two-dimensional subspace of phase space which has the property that if we
restrict the symplectic form to it, then we obtain a new symplectic form,
defined on this two-dimensional space. For instance, it is easy to check that
the $x_{j},p_{j}$ are symplectic planes (but those of coordinates $x_{j}%
,p_{k}$, $j\neq k$, or $x_{j},x_{k}$, $p_{j},p_{k}$ are not). One proves
\cite{Arnold,Birk} that every symplectic plane can be obtained from any of the
$x_{j},p_{j}$ planes using a symplectic transformation. This property implies,
in particular, that the Robertson--Schr\"{o}dinger inequalities (\ref{robup})
are covariant under symplectic transformations: if one defines new coordinates
$x^{\prime},p^{\prime}$ by $(x^{\prime},p^{\prime})^{T}=S(x,p)^{T}$, $S$ a
symplectic matrix, then if
\[
(\Delta p_{j})^{2}(\Delta x_{j})^{2}\geq\Delta(x_{j},p_{j})^{2}+\tfrac{1}%
{4}\hbar^{2}%
\]
we also have%
\[
(\Delta p_{j}^{\prime})^{2}(\Delta x_{j}^{\prime})^{2}\geq\Delta(x_{j}%
^{\prime},p_{j}^{\prime})^{2}+\tfrac{1}{4}\hbar^{2}.
\]

Also, there are possible non-trivial generalizations of the uncertainty
principle, using new results in symplectic topology, for instance \cite{Abbo}
which extends Gromov's theorem (in the linear case) to projections on
symplectic subspaces with dimension greater than $2$. In \cite{gossang} we
have shown how this result leads to \textquotedblleft quantum universal invariants".

\section{Conclusion}

Quoting the great mathematician Hermann Weyl:

\begin{quotation}
`\textit{In these days the angel of topology and the devil of abstract algebra
fight for the soul of each individual mathematical domain' (H. Weyl, 1939)}
\end{quotation}

This quotation from goes straight to the point, and applies to Physics as
well: while algebra (in the large) has dominated the scene of quantum
mechanics for a very long time (in fact, from its beginning), we are
witnessing a slow but steady emergence of geometric ideas. Not only do these
geometric ideas add clarity to many concepts, but they also lead to new
insights (see e.g. \cite{gossang}). This is what we had in mind while writing
the present paper.

\begin{acknowledgement}
The present work has been supported by the Austrian Research Agency FWF
(Projektnummer P20442-N13).
\end{acknowledgement}

\begin{acknowledgement}
I wish to express my gratitude to my son Sven for having drawn the pictures in
this paper.
\end{acknowledgement}

\end{document}